\documentclass[prb,twocolumn,superscriptaddress,showpacs,aps]{revtex4-1}

\usepackage{color}
\usepackage{graphicx}
\usepackage{bm}
\usepackage{amsmath}
\usepackage{multirow}
\usepackage{epstopdf}

\begin{document}

\newcommand{\U}{URu$_2$Si$_2$}
\newcommand{\UFe}{URu$_{2-x}$Fe$_x$Si$_2$}
\newcommand{\Tc}{T_{\text c}}

\title{Chemical pressure tuning of \U\ via isoelectronic substitution of Ru with Fe}

\author{Pinaki Das}
\affiliation{MPA-CMMS, Los Alamos National Laboratory, Los Alamos, New Mexico 87545, USA}

\author{N. Kanchanavatee}
\affiliation{Department of Physics, University of California, 됩an Diego, La Jolla, California 92093, USA}
\affiliation{Center for Advanced Nanoscience, University of California, San Diego, La Jolla, California 92093, USA}

\author{J. S. Helton}
\affiliation{The United States Naval Academy, Department of Physics, Annapolis, Maryland 21402, USA}
\affiliation{NIST Center for Neutron Research, National Institute of Standards and Technology, Gaithersburg, Maryland 20899, USA}

\author{K. Huang}
\affiliation{Center for Advanced Nanoscience, University of California, San Diego, La Jolla, California 92093, USA}
\affiliation{Materials Science and Engineering Program, University of California, San Diego, La Jolla, California 92093, USA}

\author{R. E. Baumbach}
\affiliation{MPA-CMMS, Los Alamos National Laboratory, Los Alamos, New Mexico 87545, USA}
\affiliation{National High Magnetic Field Laboratory, Florida State University, Tallahassee , Florida 32310, USA}

\author{E. D. Bauer}
\affiliation{MPA-CMMS, Los Alamos National Laboratory, Los Alamos, New Mexico 87545, USA}

\author{B. D. White}
\affiliation{Department of Physics, University of California, 됩an Diego, La Jolla, California 92093, USA}
\affiliation{Center for Advanced Nanoscience, University of California, San Diego, La Jolla, California 92093, USA}

\author{V. W. Burnett}
\affiliation{Department of Physics, University of California, 됩an Diego, La Jolla, California 92093, USA}
\affiliation{Center for Advanced Nanoscience, University of California, San Diego, La Jolla, California 92093, USA}

\author{M. B. Maple}
\affiliation{Department of Physics, University of California, 됩an Diego, La Jolla, California 92093, USA}
\affiliation{Center for Advanced Nanoscience, University of California, San Diego, La Jolla, California 92093, USA}
\affiliation{Materials Science and Engineering Program, University of California, San Diego, La Jolla, California 92093, USA}

\author{J. W. Lynn}
\affiliation{NIST Center for Neutron Research, National Institute of Standards and Technology, Gaithersburg, Maryland 20899, USA}

\author{M. Janoschek}
\email{mjanoschek@lanl.gov}
\affiliation{MPA-CMMS, Los Alamos National Laboratory, Los Alamos, New Mexico 87545, USA}

\date{\today}

\begin{abstract}
We have used specific heat and neutron diffraction measurements on single crystals of URu$_{2-x}$Fe$_x$Si$_2$ for Fe concentrations $x$~$\leq$~0.7 to establish that chemical substitution of Ru with Fe acts as ``chemical pressure'' $P_{ch}$ as previously proposed by Kanchanavatee {\it et al.}~[Phys. Rev. B {\bf 84}, 245122 (2011)] based on bulk measurements on polycrystalline samples. Notably, neutron diffraction reveals a sharp increase of the uranium magnetic moment at $x=0.1$, reminiscent of the behavior at the ``hidden order'' (HO) to large moment antiferromagnetic (LMAFM) phase transition observed at a pressure $P_x$~$\approx$~0.5-0.7~GPa in URu$_2$Si$_2$. Using the unit cell volume determined from our measurements and an isothermal compressibility $\kappa_{T} = 5.2 \times 10^{-3}$ GPa$^{-1}$ for \U, we determine the chemical pressure $P_{ch}$ in \UFe\ as a function of $x$. The resulting temperature $T$-chemical pressure $P_{ch}$ phase diagram for \UFe\ is in agreement with the established temperature $T$-external pressure $P$ phase diagram of \U.
\end{abstract}


\maketitle

\section{Introduction}

The heavy fermion compound \U\ exhibits a large entropy change at a temperature $T_0$ = 17.5~K~\cite{Palstra:85, Maple:86, Schlablitz:86} that has presented a challenge to researchers for almost 30 years. Despite the pronounced signature at $T_0$ in the heat capacity that signals a second-order symmetry breaking phase transition, the order parameter of the ground state below $T_0$ remains unknown, and the phase is commonly referred to as the ``hidden order'' (HO) phase.\cite{Luethi:93}

On the other hand, a large body of experimental work has established that the HO phase develops out of a paramagnetic phase that is characterized by strong electronic correlations stemming from the hybridization of localized uranium $f$ electrons with the conduction electrons.\cite{Mydosh:11} Scanning tunneling microscopy (STM) measurements on \U\ demonstrate that hybridization already arises at temperatures much larger than $T_0$, where the single-ion Kondo temperature was determined as $T_K$ = 120~K.\cite{Schmidt:10, Aynajian:10} At $T^\ast$~$\approx$~70 K, coherence sets in and a Kondo lattice develops as is observed by point-contact spectroscopy (PCS) measurements\cite{Rodrigo:97} consistent with bulk properties.\cite{Palstra:86, Schoenes:87} NMR measurements \cite{Shirer:13} also suggest the presence of a pseudogap below 30 K.

The onset of the HO is accompanied by a further reorganization of this strongly correlated electron state. The BCS-like specific heat anomaly at $T_0$,\cite{Palstra:85,Maple:86} Hall effect\cite{Schoenes:87, Oh:07, Kasahara:07} and optical conductivity~\cite{Bonn:88} measurements show that a charge gap opens over about 40\% of the Fermi surface (FS) within the HO state. The partial gapping of the FS is further supported by quantum oscillation measurements~\cite{Ohkuni:99, Altarawneh:11} that reveal that the FS of \U\ consists of mostly small closed pockets. This reorganization of the electronic structure below $T_0$ was also observed via angle-resolved photoemission spectroscopy (ARPES),\cite{Santander-Syro:09} STM,\cite{Schmidt:10, Aynajian:10} and PCS~\cite{Rodrigo:97} that reveal a secondary hybridization of a heavy $f$-like quasiparticle band with a light hole-like band at $Q^\ast=\pm0.3\pi/a$, resulting in the formation of a hybridization gap $\Delta_{Q^\ast}$~=~5~meV. More recent ARPES work suggests that possibly larger regions of the FS are gapped at $T_0$.\cite{Bareille:14}

Extensive inelastic neutron scattering studies demonstrated that charge and spin degrees of freedom are strongly coupled in \U.\cite{} Below $T_0$, spin gaps develop simultaneously with the charge gap, where most of the work has focused on the spin gaps at the commensurate $\bm{Q}_0 = (1, 0, 0)$ and incommensurate $\bm{Q}_1 = (0.4, 0, 0)$ wave vectors.\cite{Broholm:91, Wiebe:07, Villaume:08, Bourdarot:10, Bourdarot:14} The $\bm{Q}_1$ mode has been attributed to itinerant-like spin excitations that are related to the heavy electronic quasiparticles that form below $T^\ast$.~\cite{Wiebe:07} A more recent neutron scattering study that has investigated the magnetic scattering over large parts of the Brillouin zone shows that spin gaps open over wider regions of reciprocal space, and the observed magnetic excitations originate from spin-flip transitions between hybridized bands that track the FS\cite{Butch:14} as previously suggested based on a smaller set of data.\cite{Janik:09} The gapping of these spin fluctuations accounts for the loss of entropy at $T_0$.\cite{Wiebe:07}

The $\bm{Q}_0$ mode transforms to weak quasielastic spin fluctuations above $T_0$ and appears to be a true signature of the HO state, where the integrated dynamic spin susceptibility behaves like an order parameter.\cite{Bourdarot:10} Notably, when \U\ is tuned by means of external pressure, above a pressure $P_x$~$\approx$~0.5-0.7~GPa, the commensurate spin gap closes and the longitudinal spin fluctuations at $Q_0$ freeze out,\cite{Villaume:08} leading to the formation of antiferromagnetic order with a magnetic moment of $\sim~ 0.4~\mu_B$/U parallel to the tetragonal $c$ axis.\cite{Amitsuka:99} We note that antiferromagnetic order with the same ordering wave vector $\bm{Q}_0$ and a tiny magnetic moment $\sim~ 0.01-0.04 ~\mu_B$/U \cite{Broholm:87,Broholm:91} is also observed in the HO state for $P<P_x$. Historically, the antiferromagnetic phase beyond $P_x$ has therefore been called the large moment antiferromagnetic phase (LMAFM). However, the moment in the HO phase is too small to account for the large entropy 0.2$R$ln2 associated with the specific heat anomaly below $T_0$.\cite{Maple:86, Palstra:85} Detailed subsequent investigations led to the wide consensus that the small magnetic moment within the HO state is due to internal strain.\cite{Amitsuka:07, Niklowitz:10} Larmor diffraction measurements have additionally established that the phase boundary between the HO and LMAFM is first order.\cite{Niklowitz:10} Recent theoretical work\cite{Rau:12,Chandra:13} suggested that the magnetic moment in the HO phase has a small component in the tetragonal plane, and that the antiferromagnetic order in the HO and LMAFM phases would be different, but this was ruled out by recent detailed neutron diffraction work that confirmed  that the magnetic moment in the HO phase is purely along the $c$-axis.\cite{Das:13, Metoki:13,Ross:14}

It is therefore well-established that the two phases have distinct order parameters. In contrast, quantum oscillation measurements suggest that the Fermi surfaces of the LMAFM and HO are nearly identical.\cite{Hassinger:10} This is partially supported by inelastic neutron scattering experiments that probed the incommensurate spin excitations at $\bm{Q}_1$ as a function of pressure, which find that the FS nesting vector remains unchanged upon entering the LMAFM phase.\cite{Bourdarot:14} However, the incommensurate spin gap at $\bm{Q}_1$ is found to increase abruptly at $P_x$ from approximately 4.5 to just below 8~meV.\cite{Bourdarot:14} Complete measurements spanning the entire Brillouin zone are difficult to perform under pressure and a full comparison with the HO phase is thus not possible. This demonstrates that the relationship between the HO and LMAFM phases that is determined via a complex interplay of localized and itinerant electronic, as well as spin degrees of freedom, remains an open question. In particular, it would be desirable to perform measurements such as STM and ARPES that are able to detect the fine details of the electronic structure in the LMAFM phase to resolve these questions; these experiments have not yet been performed since these measurements cannot be carried out under applied pressure.

Here we demonstrate that the new isoelectronic substitution series {\UFe} may represent a new route to investigate the interplay between HO and LMAFM phases. Measurements of electrical resistivity, magnetic susceptibility and specific heat on polycrystalline samples of \UFe\ for $0 ~\le~ x ~\le~ 2$ have recently been reported; from these measurements, a phase diagram as a function of Fe concentration $x$ that tracks the $T-P$ phase diagram of \U was established.~\cite{Kanchanavantee:11} Notably, it appears that the reduction in the unit cell volume arising from the the substitution of the smaller isoelectronic Fe ions for Ru acts as ``chemical pressure". However, it is difficult to determine the phase boundary between the HO and LMAFM phases from bulk measurements, because they do not directly probe the order parameter of the LMAFM phase. Therefore, it was impossible to determine whether \UFe\ samples with $x$~$\gtrsim$~0.2 indeed were in the LMAFM ground state. In addition, the polycrystalline samples also exhibited some issues with disorder, which led to broadened phase transitions as a function of temperature for $x$~$>$~0.1, thus leading to some uncertainty in the phase diagram. In the work presented here, we have overcome these issues by performing specific heat and neutron diffraction experiments on high-quality single crystals of \UFe\ for $0 ~\le~ x ~\le~ 0.7$.  The results are consistent with the viewpoint that substitution of Fe for Ru in \UFe\ acts as a chemical pressure and, up to at least $x$~$=$~0.3, reproduces the temperature $T$-external pressure $P$ phase diagram of \U.

\section{Experimental Methods}

Measurements were performed on a series of single crystals of {\UFe} with Fe concentrations $0.025 \le x \le 0.7$. Samples were grown using the Czochralski technique, where one of the samples ($x=0.1$) was prepared in a tri-arc furnace with a continuously purified Ar atmosphere at Los Alamos, and all other samples were prepared in a Techno Search TCA 4-5 Tetra-Arc furnace under a zirconium-gettered argon atmosphere at UCSD. The quality of the synthesized single crystals was confirmed by x-ray diffraction measurements in a D8 Discover Bruker diffractometer. Typical sample masses for all samples were a few grams.  Specific heat measurements were performed for 1.8 K $\le T \le 30$ - 50~K in a Quantum Design Physical Property Measurement System semiadiabatic calorimeter using a heat-pulse technique. Neutron diffraction experiments were performed at the NIST Center for Neutron Research on the BT4 and BT7 triple axis spectrometers with incident energies of $E_i = 35, 75, 80$ or $81.8$ meV and various collimations.\cite{Lynn:12} Samples were oriented in the $[HK0]$ scattering plane. For $E_i$ = 35 meV, pyrolytic graphite (PG) filters were used to reduce higher-order contamination from the monochromated neutron beam. Temperatures in the range $4 \le T \le 30$ K were accessed using a closed cycle $^4$He refrigerator. The higher incident neutron energies have been used to avoid extinction. The use of the analyzer crystal on both instruments further improved the signal-to-noise ratio, which is particularly important in the case of measuring the small magnetic moments in the HO phase.

\section{Experimental Results}

\begin{figure}
  \includegraphics{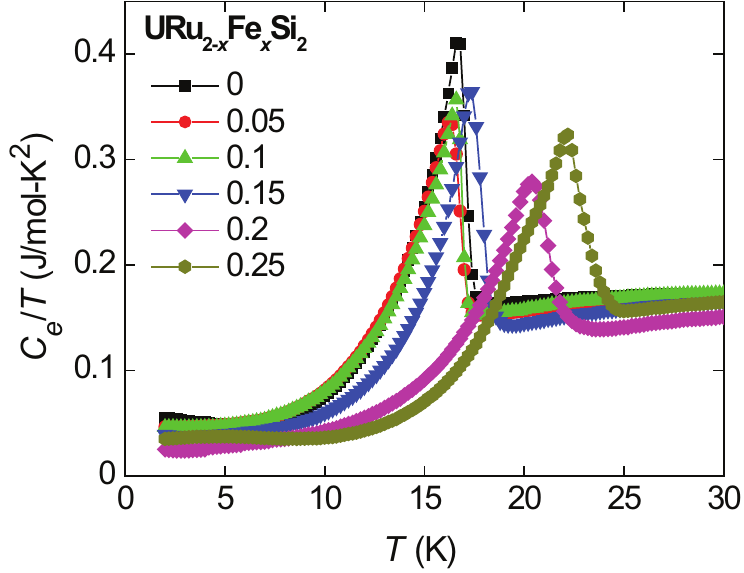}
  \caption{\label{fig1}
           (color online) Electronic specific heat $C_e$ divided by temperature $T$ vs. $T$ for single crystals of \UFe. The phonon contribution to the specific heat was subtracted according to the procedure described in the text.}
\end{figure}

In Fig.~\ref{fig1}, we show the electronic contribution to the specific heat $C_e(T)$ divided by temperature $T$ that was determined for $0 \le x \le 0.25$ by subtracting the phonon contribution $C_{ph}(T)$.\cite{Kanchanavatee:15} Specific heat data measured on the isostructural compound UFe$_2$Si$_2$ were used to estimate $C_{ph}(T)$. This method should yield a good estimate of the phonon contribution for all values of $x$ since the end member compounds are isostructural and UFe$_2$Si$_2$ is reported to be a Pauli paramagnet down to 0.2 K.\cite{Szytula:88} Using only a Debye function, we were unable to account correctly for the phonon contribution over the entire $T$-range measured. $C_e(T)$ shows a well-defined BCS-like anomaly at the transition from the paramagnetic phase into the HO/LMAFM phase at $T_0$ for all measured $x$ that is similar to the feature observed for $x=0$.\cite{Maple:86} We defined $T_0$ as the temperature of the midpoint of this feature in the $C_e(T)$ data. There is no sign of the broad shoulder above $T_0$ that was observed for polycrystalline samples previously and has been attributed to disorder,~\cite{Kanchanavantee:11} demonstrating that the single crystals studied here do not suffer from similar problems. For increasing Fe concentration, $x$, the transition temperature $T_0$ moves to higher temperatures as observed for \U\ under applied pressure.\cite{}

\begin{figure}
  \includegraphics{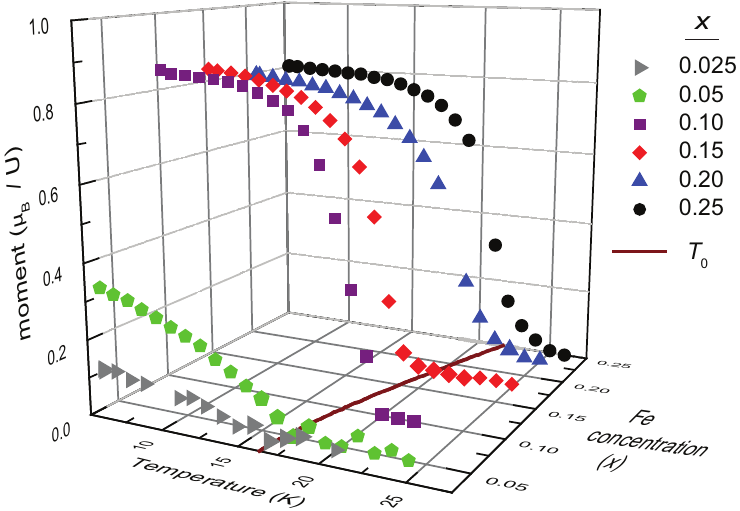}
  \caption{\label{3Dplot}
           (color online) Temperature dependence $T$ of the magnetic moment per uranium ion in \UFe\ for various Fe concentrations $x$ determined via magnetic neutron diffraction (see text for details). The solid (brown) line defines the transition temperatures $T_0$ from the paramagnetic state into the hidden order (HO)/large moment antiferromagnetic (LMAFM) phase as extracted from order parameter fits (see text for details).}
\end{figure}

Figure~\ref{3Dplot} shows the magnetic moment per uranium as a function of temperature $T$ for all measured Fe concentrations $x$ obtained from our neutron scattering experiments. The magnetic moment for each temperature and Fe concentration was obtained by recording the integrated intensity of the magnetic Bragg reflection (1,0,0) by means of rocking scans. We note that (1,0,0) is a forbidden nuclear Bragg reflection and the observed scattering is purely magnetic apart from higher order scattering from the monochromator that was temperature independent. The intensity of a magnetic reflection for \UFe\ is given by

\begin{equation}
I_{\rm M}(\bm{Q})=(\gamma r_0)^2\frac{(2\pi)^3}{v_0}\sum_{\alpha,\beta}(\delta_{\alpha\beta}-\hat{Q}_{\alpha}\hat{Q}_{\beta})F_M(\bm{Q})^{\alpha\dagger}F_M(\bm{Q})^{\beta},
\end{equation}
where $\gamma = 1.193$ is the magnetic dipole moment of the neutron in units of nuclear Bohr magnetons, $r_0 = 2.818 \times 10^{-15}$ m, and $v_0$ is the unit cell volume. $\hat{Q}$ is a unit vector parallel to the scattering vector $\bm{Q}$ and $\bm{F}_M(\bm{Q})$ is the magnetic structure factor. The indices $\alpha$ and $\beta$ describe their components with $\alpha,\beta = x,y,z$. The term,  $\sum_{\alpha,\beta}(\delta_{\alpha\beta}-\hat{Q}_{\alpha}\hat{Q}_{\beta})$, signifies that only components of the magnetic structure factor perpendicular to $\bm{Q}$ contribute to the magnetic scattering. The magnetic structure factor for \UFe\ is described by
\begin{eqnarray}
\bm{F}_M(\bm{Q})&=&g F_{\bm{Q}}\exp(-W_{\bm{Q}})\sum_d\exp(i\bm{Q}\cdot\bm{d})\bm{S}_d\nonumber\\
                &=&g F_{\bm{Q}}\exp(-W_{\bm{Q}})\left[\left(\begin{array}{c}0\\0\\S\end{array}\right)-\left(\begin{array}{c}0\\0\\-S\end{array}\right)\right].
\end{eqnarray}
Here $g$, $F_{\bm{Q}}$ and $\exp(-W_{\bm{Q}})$ are the Land\'{e} $g$-factor, the magnetic form factor and the Debye-Waller factor of the magnetic uranium ions, respectively. The vector $\bm{d}$ describes the position of the $d$th uranium ion in the unit cell. In the second line, we have used the fact that \UFe\ crystallizes in the space group $I 4/m m m$, and the antiferromagnetic order found in the HO/LMAFM state is described by the two uranium ions (0,0,0) and (1/2, 1/2, 1/2) that exhibit antiparallel magnetic moments $\bm{S}_{(0,0,0)} = (0,0,S)$ and $\bm{S}_{(1/2, 1/2, 1/2)} = (0,0,-S)$. The magnetic moment per uranium is therefore given by $m = gS$ and can be calculated by normalizing the recorded magnetic intensity with the known intensity of nuclear reflections such as (2$n$,0,0). We note that we have used weak nuclear reflection such as (6,0,0) and higher incident neutron energies $E_i$ to avoid errors in the normalization due to extinction effects.

\section{Discussion}

\begin{figure}
  \includegraphics[scale=0.8]{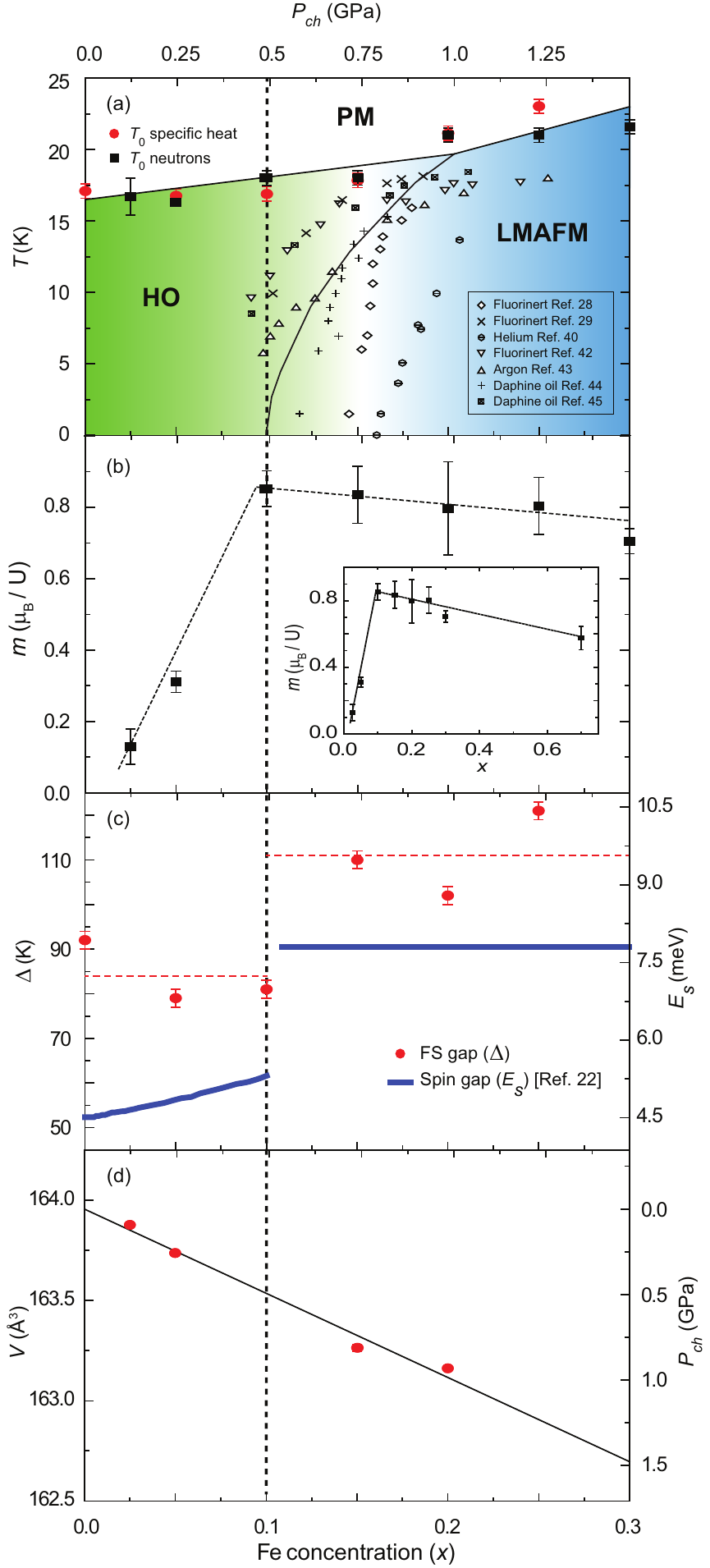}
  \caption{\label{phasediagram}
           (color online) Summary of the main results of our study on single crystals of \UFe. (a) Temperature $T$ vs. Fe concentration $x$ phase diagram constructed from specific heat and neutron diffraction measurements. The Fe concentration $x$ was converted to ``chemical pressure'' $P_{ch}$ on the top horizontal axis (see text). We have also plotted the phase boundaries in \U\ obtained by various pressure studies for comparison.\cite{Bourdarot:05,Hassinger:08,Amitsuka:07,Amitsuka:08,Niklowitz:10,Motoyama:08,Butch:10} The acronyms PM, HO and LMAFM denote the paramagnetic, hidden order and large moment antiferromagnetic phases, respectively. (b) The magnetic moment $m$ per uranium site vs. $x$ at $T$~$=$~5~K determined from our neutron diffraction measurements (see text). The inset shows the full range of measured $x$ up to 0.7. (c) Energy gap $\Delta$ that opens on the FS due to the onset of the HO phase as determined via fits to the low temperature electronic specific heat $C_e(T)$ data (see text). We also show the size of the incommensurate spin gap $E_S$ at the wave vector $\bm{Q}_1 = (0.4, 0, 0)$, determined via inelastic neutron scattering, vs. $P_{ch}$ for comparison.\cite{Bourdarot:10} (d) Unit cell volume $V$ vs. $x$. For the axis on the right side (and the top horizontal axis), the unit cell volume was converted to chemical pressure (see text). The lines are guides to the eye.}
\end{figure}

From Fig.~\ref{3Dplot}, it is clearly visible that the onset temperature $T_0$ of the HO/LMAFM phase increases for increasing $x$. Here, we have determined $T_0(x)$ from the temperature dependence of the magnetic moment for each concentration by performing order parameter fits of the form $m \propto (T_{0}-T)^{\beta}$.\cite{Butch:10} The values of $T_0(x)$ extracted in this fashion are marked with the brown solid line in Fig.~\ref{3Dplot} and are also shown in the temperature $T$-Fe concentration $x$ phase diagram in Fig.~\ref{phasediagram}(a) together with the values obtained from the specific heat measurements. For the latter, $T_0$ was obtained by tracking the peak of the jump in $C_e(T)/T$. Figure~\ref{phasediagram}(a) illustrates that the values of $T_0(x)$ derived from both measurements are consistent. The shape of $T_0(x)$ looks remarkably similar to the shape of the curve $T_0(P)$ observed when \U\ is tuned via external pressure $P$. Notably, we observe a kink in $T_0(x)$ at $x$~$\approx$~0.15 that occurs in $T_0(P)$ at $P_{\rm kink}$~$\approx$~0.7-1.3 GPa depending on the combination of method and pressure medium employed to determine $T_0(P)$.\cite{Bourdarot:05,Hassinger:08,Amitsuka:07,Amitsuka:08,Niklowitz:10,Motoyama:08,Butch:10}

In the $T$-$P$ phase diagram of \U, the HO-LMAFM phase transition for $T\rightarrow$~0~K occurs at $P_x$~$\approx$~0.5-0.7 GPa, where the phase boundary moves to slightly higher pressures for increasing temperature, until it connects to the phase boundary between the paramagnetic phase and the HO/LMAFM phase $T_0(P)$ at $P_{\rm kink}$ (see Fig.~\ref{phasediagram}(a)).\cite{Bourdarot:05,Hassinger:08,Amitsuka:07,Amitsuka:08,Niklowitz:10,Motoyama:08,Butch:10} In Fig.~\ref{phasediagram}(b), we show the value of the U magnetic moment $m$ as a function of $x$ at $T$~=~5~K. The order parameter curves $m(T)$ illustrated in Fig.~\ref{3Dplot} for various values of $x$ demonstrate that the magnetic moment is already saturated at $T$~=~5~K, so that the value of $m$ plotted in Fig.~\ref{phasediagram}(b) effectively corresponds to the magnetic moment for $T\rightarrow$~0~K. The magnetic moment only increases slightly as a function of $x$ for $x$~$<$~0.1; however, at $x$~$=$~0.1, a sharp increase of $m$ is observed, indicating that $x$~$\approx$~0.1 marks the transition from the HO to the LMAFM phase in \UFe\ for $T\rightarrow$~0~K.

This is also apparent from a more detailed analysis of the specific heat data, which allows us to investigate the magnitude of the charge gap $\Delta$ that opens in the FS at $T_0$ as a function of Fe concentration. The specific heat $C_{e}(T)$ data below $T_0$ can be well described by the expression\cite{Maple:86}
\begin{equation}\label{DeltaFit}
  C_{e}(T)= A\exp(-\Delta/T).
\end{equation}
The values of $\Delta$ extracted by fitting Eq.~\ref{DeltaFit} to the electronic contribution to the specific heat for each $x$ are plotted in Fig.~\ref{phasediagram}(c). The FS gap $\Delta$~$\approx$ $85$ K remains roughly constant up to $x = 0.1$, where a sudden increase to approximately $\sim 110$ K is observed, highlighting a change in the electronic structure of \UFe\ at the same Fe concentration, where the abrupt change of the magnetic moment indicates the phase transition from the HO into the LMAFM state.

We note that the behavior of the charge gap $\Delta$ extracted from the specific heat data is reminiscent of the sudden increase of the spin gap $E_S$ at the incommensurate wave vector $\bm{Q}_1$ when \U\ is tuned from the HO into the LMAFM phase by means of external pressure.\cite{Villaume:08,Bourdarot:10} We have highlighted this similarity by overlaying the values of the spin gap $E_S$ as a function of pressure extracted from inelastic neutron scattering measurements in Fig.~\ref{phasediagram}(c). Because the spin excitations at $\bm{Q}_1$ represent spin-flip transitions between hybridized bands and track the gapped FS of \U,\cite{Butch:14,Janik:09} it is natural that the spin gap $E_S$ observed via neutron scattering tracks the charge gap $\Delta$ probed by the specific heat.

The abrupt changes of the magnetic moment $m$ and the charge gap $\Delta$ at $x$ = 0.1, as well as the shape of the phase boundary $T_0(x)$, support the interpretation that isoelectronic substitution of Ru with smaller Fe ions in \U\ behaves as chemical pressure. In Fig.~\ref{phasediagram}(d), we show the unit cell volume $V$ of \UFe\ as a function of $x$ as extracted from x-ray diffraction measurements. As expected, and similar to the previously investigated polycrystalline samples,\cite{Kanchanavantee:11} $V$ decreases for increasing $x$. Using the value for the isothermal compressibility ($\kappa_{T} = 5.2 \times 10^{-3}$ GPa$^{-1}$) of \U\ as in our previous work \cite{Kanchanavantee:11} and as determined from x-ray diffraction at low temperature,\cite{Kuwahara:03} we calculate the chemical pressure $P_{ch}$ that corresponds to each $x$ (see right side axis of Fig.~\ref{phasediagram}(d)). However, one should note that values of $\kappa_{T}$ reported in the literature range from $2 \times 10^{-3}$ to $7.3 \times 10^{-3}$ GPa$^{-1}$.\cite{Amitsuka:99,Kuwahara:03}

We also indicate $P_{ch}$ on the top horizontal axis of Fig.~\ref{phasediagram}. Using the estimated chemical pressure, we have plotted the phase boundaries between the paramagnetic, HO and LMAFM phases for \U\ as a function of pressure, as reported by several groups using different methods and pressure media.\cite{Bourdarot:05,Hassinger:08,Amitsuka:07,Amitsuka:08,Niklowitz:10,Motoyama:08,Butch:10} It is immediately apparent that the phase boundaries determined here for \UFe\ and plotted as a function of chemical pressure are in good agreement with the established $T$-$P$ phase diagram for \U.

We note that the magnetic moment within the HO phase of \UFe\ increases faster than observed for pressure tuning of \U. It is likely that this discrepancy is due to the additional disorder that is introduced by chemical substitution. In this case, the disorder introduces additional strain in \UFe\ for increasing $x$, which leads to larger ``puddles" of the LMAFM phase existing within the HO state. The magnetic moment $m$~$\approx$~0.8 $\mu_B$/U in the LMAFM phase of \UFe\ is slightly larger than what has been observed in the LMAFM phase accessed via external pressure ($m$~=~0.4-0.5 $\mu_B$/U).\cite{Amitsuka:07, Butch:10} This may also be related to the disorder introduced by chemical substitution and/or an increased magnetic moment due to the introduction of Fe; however, additional more detailed studies will be required to explore these subtle differences between external and chemical pressure tuning.

Next, we discuss the slight difference in Fe concentration $x$ at which the HO-LMAFM phase transition occurs when we compare previous bulk measurements on polycrystalline samples of \UFe\ ($x$~$\approx$~0.2) with the value determined for single crystals herein ($x$~$\approx$~0.1). First, we note that the value for the polycrystalline samples was based on the position of the kink in the phase boundary $T_0(x)$ and an analysis of the entropy, which are both not directly sensitive to the order parameter of the LMAFM phase. In contrast, the value determined here is based on the change of the magnetic moment on the U site as $T\rightarrow$~0~K. As described above, the phase boundary between the HO and LMAFM phases moves to higher pressure/Fe concentration as a function of increasing temperature. Furthermore, we note that the polycrystalline samples show some disorder as apparent from a broadened HO/LMAFM transition, which makes it more difficult to determine the exact shape of the $T_0(x)$ phase boundary. Taken together, these factors account for the slight difference between poly- and single-crystalline samples.

Recently, the HO/LMAFM phase boundary $T_0(x)$ was determined for the system URu$_{2-x}$Os$_x$Si$_2$ to values of $x$ = 1.2, based on electrical resistivity, magnetic susceptibility, and specific heat measurements on polycrystalline specimens.\cite{Kanchanavatee14,Dalichaouch90}  Similar to the URu$_{2-x}$Fe$_x$Si$_2$ system, $T_0$ for the URu$_{2-x}$Os$_x$Si$_2$ system increases with $x$ and exhibits a maximum of $\sim$50 K at $x$ = 1.0 (somewhat larger than the maximum of $\sim$42 K for the Fe-substituted system at $x$ = 0.8).  Since the substitution of isoelectronic Os for Ru increases the unit cell volume with $x$, it should act as a negative chemical pressure and result in a decrease of $T_0$ with $x$, in analogy with the apparent equivalence of chemical pressure in the behavior of $T_0(x)$ of URu$_{2-x}$Fe$_x$Si$_2$ and applied pressure in the behavior of $T_0(P)$ of URu$_2$Si$_2$. Thus, the similarity in the behavior of $T_0(x)$ for the isoelectronic Os and Fe substitutions for Ru in URu$_2$Si$_2$ is a surprising result that suggests that other factors may be involved in the behavior of $T_0(x)$, at least for the Os-substituted system.  As suggested in Ref.~\onlinecite{Kanchanavatee14}, the similar trends for $T_0$ observed for Fe and Os substitutions in \U\ correlate with similar behavior of the ratio of the lattice parameters $c/a$, which increase with $x$ for both systems. Further experiments are underway on the URu$_{2-x}$Os$_x$Si$_2$ system to address this issue.

\section{Concluding Remarks}

In summary, the $T$-$P_{ch}$ phase diagram, as well as the magnetic moment on the uranium site $m$ as a function of $P_{ch}$ that we have determined for \UFe\ from the specific heat and magnetic neutron diffraction measurements presented herein, are both in good agreement with the $T$-$P$ phase diagram and the evolution of the magnetic moment as a function of $P$ in \U. This suggests that substituting smaller Fe ions for Ru in \U\ is equivalent to applying external pressure as was proposed in earlier work.\cite{Kanchanavantee:11} The substitution series \UFe\, thus enables us to study the effects of pressure tuning on \U\ with methods such as ARPES or STM that cannot be employed under pressure. This will provide a new opportunity to study the change of the electronic structure between the HO and LMAFM phases in \U\ that are known to be intimately connected. Keeping in mind that electronic and spin degrees of freedom are closely coupled in \U, it would be desirable to probe the complete spin excitation spectrum within the Brillouin zone for an Fe concentration $x$~$\geq$~0.1 to determine subtle differences between both phases. Ultimately, these future experiments on the nature of the LMAFM phase may allow us to obtain a fresh view on the elusive order parameter of the HO state.

\section{Acknowledgements}

The research at UCSD was supported by the U.S. Department of Energy, Office of Basic Energy Sciences, Division of Materials Sciences and Engineering under Grant No. DE-FG02-04ER46105 (sample synthesis) and the National Science Foundation under Grant No. DMR-0802478 (sample characterization).  Work at Los Alamos National Laboratory (LANL) was performed under the auspices of the U.S. DOE, OBES, Division of Materials Sciences and Engineering. The identification of any commercial product or trade name does not imply endorsement or recommendation by NIST. We thank William Ratcliff and Yang Zhao for technical support during the experiments.



\begin{thebibliography}{99}


\bibitem{Palstra:85}
T.~T.~M. Palstra, A.~A. Menovsky, J.~van~den Berg, A.~J. Dirkmaat, P.~H. Kes, G.~J. Nieuwenhuys, and J.~A. Mydosh, Phys. Rev. Lett. {\bf 55},  2727  (1985).

\bibitem{Maple:86}
M.~B. Maple, J.~W. Chen, Y. Dalichaouch, T. Kohara, C. Rossel, M.~S. Torikachvili, M.~W. McElfresh, and J.~D. Thompson, Phys. Rev. Lett. {\bf 56},  185  (1986).

\bibitem{Schlablitz:86}
W. Schlabitz, J. Baumann, B. Pollit, U. Rauchschwalbe, H.~M. Mayer, U. Ahlheim, and C.~D. Bredl, Z. Phys. B: Condens. Matter {\bf 62},  171  (1986).

\bibitem{Luethi:93}
 B. Luethi, B. Wolf, P. Thalmeier, M. Gunther, W. Sixl, and G. Bruls, Phys. Lett. A {\bf 175} 237-240 (1993).

\bibitem{Mydosh:11}
J. A. Mydosh, and P. M. Oppeneer, Rev. of Mod. Phys. {\bf 83}, 1301 (2011).

\bibitem{Schmidt:10}
A.~R. Schmidt, M.~H. Hamidian, P. Wahl, F. Meier, A.~V. Balatsky, J.~D. Garrett, T.~J. Williams, G.~M. Luke, and J.~C. Davis, Nature {\bf 465},  570  (2010).

\bibitem{Aynajian:10}
P. Aynajian, E.~H. da Silva Neto, C.~V. Parker, Y. Huang, A. Pasupathy, J. Mydosh, and A. Yazdani, PNAS {\bf 107},  10383 (2010).

\bibitem{Rodrigo:97}
J.~G. Rodrigo, F. Guinea, S. Vieira, and F.~G. Aliev, Phys. Rev. B {\bf 55}, 14318 (1997).

\bibitem{Palstra:86}
T. T. M. Palstra, A. A. Menovsky, and J. A. Mydosh, Phys. Rev. B 33, 6527(R) (1986).

\bibitem{Schoenes:87}
J. Schoenes, C. Schonenberger, J. J. M. Franse, and A. A. Menovsky, Phys. Rev. B(R) {\bf 35} 5375 (1987)

\bibitem{Shirer:13}
K. R. Shirer, J. T. Haraldsen, A. P. Dioguardi, J. Crocker, N.~ apRoberts-Warren, A. C. Shockley, C. H. Lin, D. M. Nisson, J. C. Cooley, M. Janoschek, K. Huang, N. Kanchanavatee, M. B. Maple, M. J. Graf, A. V. Balatsky, and N. J. Curro, Phys. Rev. B 88, 094436 (2013).

\bibitem{Oh:07}
Y. S. Oh, K. H. Kim, P. A. Sharma, N. Harrison, H. Amitsuka, and J. A. Mydosh, Phys. Rev. Lett. {\bf 98}, 016401 (2007).

\bibitem{Kasahara:07}
Y. Kasahara, T. Iwasawa, H. Shishido, T. Shibauchi, K. Behnia, Y. Haga, T. D. Matsuda, Y. Onuki, M. Sigrist, and Y. Matsuda, Phys. Rev. Lett. {\bf 99}, 116402 (2007).

\bibitem{Bonn:88}
D. A. Bonn, J. D. Garrett, and T. Timusk, Phys. Rev. Lett. {\bf 61} 1305 (1988)

\bibitem{Ohkuni:99}
H. Ohkuni, Y. Tokiwa, K. Sakurai, R. Settai, T. Haga, E. Yamamoto, Y. Onuki, H. Yamagami, S. Takahashi, and T. Yanagisawa, Philos. Mag. B {\bf 79} 1045 (1999).

\bibitem{Altarawneh:11}
M. M. Altarawneh, N. Harrison, S. E. Sebastian, L. Balicas, P. H. Tobash, J. D. Thompson, F. Ronning, and E. D. Bauer, Phys. Rev. Let. {\bf 106} 146403 (2011).

\bibitem{Santander-Syro:09}
A.~F. Santander-Syro, M. Klein, F.~L. Boariu, A. Nuber, P. Lejay, and F. Reinert, Nat. Phys. {\bf 5},  637  (2009).

\bibitem{Bareille:14}
C. Bareille, F. L. Boariu, H. Schwab, P. Lejay,	F. Reinert, and A. F. Santander-Syro, Nature Communications {\bf 5}, 4326 (2014).

\bibitem{Broholm:91}
C. Broholm, H. Lin, P. T. Matthews, T. E. Mason, W. J. L. Buyers, M. F. Collins, A. A. Menovsky, J. A. Mydosh, and J. K. Kjems, Phys. Rev. B {\bf 43}, 12 809 (1991).

\bibitem{Wiebe:07}
C. R. Wiebe, J. A. Janik, G. J. MacDougall, G. M. Luke, J. D. Garrett, H. D. Zhou, Y. J. Jo, L. Balicas, Y. Qiu, J. R. D. Copley, Z. Yamani, and W. J. L. Buyers, Nature Phys. {\bf 3}, 96 (2007).

\bibitem{Villaume:08}
A. Villaume, F. Bourdarot, E. Hassinger, S. Raymond, V. Taufour, D. Aoki, and J. Flouquet, Phys. Rev. B 78, 012504 (2008).

\bibitem{Bourdarot:10}
F. Bourdarot, E. Hassinger, S. Raymond, D. Aoki, V. Taufour, L.-P. Regnault, and J. Flouquet, J. Phys. Soc. Jpn., {\bf 79}, 064719 (2010).

\bibitem{Bourdarot:14}
F. Bourdarot, S. Raymond, and L.-P. Regnault, Philos. Mag., DOI: 10.1080/14786435.2014.935513 (2014).

\bibitem{Butch:14}
N. P. Butch, M. E. Manley, J. R. Jeffries, M. Janoschek, K. Huang, M. B. Maple, A. H. Said, B. M. Leu, and J. W. Lynn, accepted for publication in Phys. Rev B.

\bibitem{Janik:09}
J. A. Janik, H. D. Zhou, Y.-J. Jo, L. Balicas, G. J. Mac-Dougall, G. M. Luke, J. D. Garrett, K. J. McClellan, E. D. Bauer, J. L. Sarrao, Y. Qiu, J. R. D. Copley, Z. Yamani, W. J. L. Buyers, and C. R. Wiebe, J. Phys. Condens. Matter 21, 192202 (2009).

\bibitem{Amitsuka:99}\label{Amitsuka:99}
H. Amitsuka, M. Sato, N. Metoki, M. Yokoyama, K. Kuwahara, T. Sakakibara, H. Morimoto, S. Kawarazaki, Y. Miyako, and J. A. Mydosh, Phys. Rev. Lett. {\bf 83}, 5114 (1999).

\bibitem{Broholm:87}
C. Broholm, J.~K. Kjems, W.~J.~L. Buyers, P. Matthews, T.~T.~M. Palstra, A.~A. Menovsky and J.~A. Mydosh, Phys. Rev. Lett. {\bf 58}, 1467 (1987).

\bibitem{Amitsuka:07}
H. Amitsuka, K. Matsuda, I. Kawasaki, K. Tenya, M. Yokoyama, C. Sekine, N. Tateiwa, T. C. Kobayashi, S. Kawarazaki, and H. Yoshizawa, J. Mag. Mag. Mater. {\bf 310} 214 (2007)

\bibitem{Niklowitz:10}
P.~G. Niklowitz, C. Pfleiderer, T. Keller, M. Vojta, Y.~-K. Huang, and J.~A. Mydosh, Phys. Rev. Lett. {\bf 104},  106406 (2010).

\bibitem{Rau:12}
J. G. Rau and H. Y. Kee, Phys. Rev. B {\bf 85} 245112 (2012).

\bibitem{Chandra:13}
P. Chandra, P. Coleman, and R. Flint, Nature {\bf 493} 621 (2013).

\bibitem{Das:13}
P. Das, R. E. Baumbach, K. Huang, M. B. Maple, Y. Zhao, J. S. Helton, J. W. Lynn, E. D. Bauer, and M. Janoschek, New J. Phys {\bf 15}, 053031 (2013).

\bibitem{Metoki:13}
N. Metoki, H. Sakai, E. Yamamoto, N. Tateiwa, T. Matsuda, and Y. Haga, J Phys. Soc. Jpn. {\bf 82}, 055004 (2013).

\bibitem{Ross:14}
K. A. Ross, L. Harringer, Z. Yamani, W. J. L. Buyers, J. D. Garrett, A. A. Menovsky, J. A. Mydosh, and C. L. Broholm, Phys. Rev. B {\bf 89}, 155122 (2014).

\bibitem{Hassinger:10}
E. Hassinger, G. Knebel, T. D. Matsuda, D. Aoki, V. Taufour, and J. Flouquet, Phys. Rev. Lett. {\bf 105}, 216409 (2010).

\bibitem{Kanchanavantee:11}
N. Kanchanavatee, M. Janoschek, R. E. Baumbach, J. J. Hamlin, D. A. Zocco, K. Huang, and M. B. Maple, Phys. Rev. B {\bf 84}, 245122 (2011).

\bibitem{Lynn:12}
J. W. Lynn, Y. Chen, S. Chang, Y. Zhao, S. Chi, W. Ratcliff, B. G. Ueland, and R. W. Erwin, J. Research NIST {\bf 117}, 61 (2012).

\bibitem{Kanchanavatee:15}
N. Kanchanavatee, B. D. White, M. Janoschek, M. B. Maple (to be published).

\bibitem{Szytula:88}
A. Szytula, M. Slaski, B. Dunlap, Z. Sungaila, and A. Umezawa, J. Magn. Magn. Mater. {\bf 75}, 71 (1988).

\bibitem{Butch:10}
N. P. Butch, J. R. Jeffries, S. Chi, J. B. Le\~{a}o, J. W. Lynn, and M. B. Maple, Phys. Rev. B {\bf 82}, 060408(R) (2010).

\bibitem{Kuwahara:03}
K. Kuwahara, H. Sagayama, K. Iwasa, M. Kohgi, S. Miyazaki, J. Nozaki, J. Nogami, M. Yokoyama, H. Amitsuka, H. Nakao, and Y. Murakami, Acta Phys. Pol. B {\bf 34}, 4307 (2003).

\bibitem{Bourdarot:05}
F. Bourdarot, A. Bombardi, P. Burlet, M. Enderle, J. Flouquet, P. Lejay, N. Kernavanois, V. P. Mineev, L. Paolasini, M. E. Zhitomirsky, and B. F\r{a}k, Physica B {\bf 359-361}, 986 (2005).

\bibitem{Hassinger:08}
E. Hassinger, G. Knebel, K. Izawa, P. Lejay, B. Salce, and J. Flouquet, Phys. Rev. B {\bf 77}, 115117 (2008).

\bibitem{Amitsuka:08}
H. Amitsuka, K. Matsuda, M. Yokoyama, I. Kawasaki, S. Takayama, Y. Ishihara, K. Tenya, N. Tateiwa, T. C. Kobayashi, and H. Yoshizawa, Physica B {\bf 403}, 925 (2008).

\bibitem{Motoyama:08}
G. Motoyama, N. Yokoyama, A. Sumiyama, and Y. Oda, J. Phys. Soc. Jpn. {\bf 77}, 123710 (2008).

\bibitem{Kanchanavatee14}
N. Kanchanavatee, B. D. White, V. W. Burnett, and M. B. Maple, Philos. Mag. {\bf 94}, 3681 (2014).

\bibitem{Dalichaouch90}
Y. Dalichaouch, M. B. Maple, J. W. Chen, T. Kohara, C. Rossel, M. S. Torikachvili, and A. L. Giorgi, Phys. Rev. B {\bf 41}, 1829 (1990).

\end{thebibliography}
\end{document}